# Steady Motion of a Rigid Disk of Finite Thickness on a Horizontal Plane

Milan Batista

University of Ljubljana, Faculty of Maritime Studies and Transportation

Pot pomorscakov 4, 6320 Portoroz, Slovenia, EU

milan.batista@fpp.edu

**Abstract**

The article discusses the steady motion of a rigid disk of finite thickness rolling on its edge on a horizontal plane under the influence of gravity. The governing equations are presented and two cases allowing for a steady state solution are considered: rolling on consistently rough ground and rolling on perfectly smooth ground. The conditions of steady motion are derived for both kinds of ground and it is shown that the possible steady motion of a disk is either on a straight line in a circle. Also oscillations about steady state are discussed and conditions for stable motion are established.

## 1  Introduction

The problem of an infinitely thin rolling disk or a hoop is the classical example of non-holonomic constrained motion of a rigid body and has a long history. By its nature the problem is a special case of the motion of a rigid body of revolution on the plane for which the equation of motion was already known in the second half of the 19 century ([14]). The problem of integrating the equation of motion of a disk on rough ground was solved by P.Appel and D.Korteweg in 1900 (see [1],[7],[18]) when they showed that the first integrals of equations of motion can be expressed in terms of Gauss hypergeometric functions. Later, in 1903, E.Gallop noted that the solution of the problem of a rolling disk leads to Legendere's equation. This solution can be found in Routh's book ([15]) in which steady state motion is also considered and the formulas for the condition for steady state motion, the period of small oscillations about steady motion and the condition for stability of steady motion for the case of straight line rolling, spinning about a diameter and rolling in a circle are given. These formulas was later derived by various authors ([7],[8],[11],[12],[16],[17]). Their derivation differs from the principle from which equations of motion were derived, either from Euler equations or from Lagrange equations, and the orientation of intermediate coordinate systems used. The study of stability of motion of a thin disk on smooth ground was made for example by Milne ([9]) and recently by McDonald ([8]).

At the end of the last century the interest in the study of the motion of a disk increased. O'Reilly ([13]) made a complete study of bifurcations and stability of steady motion of a thin disk on rough and smooth ground based on two integration parameters of an analytical solution of equations of motion where, for the case of rough ground, the solution was given in terms of Legendere functions. Similar results were later given by Kuleshev ([4],[5]), who studied the rolling of a thin disk on rough ground and provided results in terms of Gauss hypergeometric functions. The solution of the problem in terms of these functions was also given in the paper of Borisov et al [2] where the bifurcation diagrams based on integration parameters were constructed and qualitative analysis of the point of contact of a disk on a plane was given.

Among papers, motivated by the paper of Moffatt ([10]) who considered the settling of a science toy called 'Euler's disk', of particular interest to this paper is a study by Kessler and O'Reilly ([3]), for the authors discuss a disk of finite thickness; they introduce the resistance moment into equations of motion and perform numerical simulations of the so-called stick-slip movement of a disk. Recently LeSaux et al ([6]) published the extensive numerical and analytical study of a thin disk rolling under various kinds of friction models.

The above review of the literature indicates that the motion of a thin disk has been, analytically and numerically, exhaustively studied. For a disk with finite thickness, as was noted by Routh ([15]), the equation of motion can be derived from the general equation of motion of the body of revolution by putting the radius of curvature at zero. This was done by Webster ([18]). But apparently, beside the Kessler and O'Reilly ([3]) numerical simulations of the motion of a disk, no study of steady motion was done. The purpose of this paper is to fill this gap.

## 2 Equations

Consider a rigid homogeneous disk moving on a horizontal plane in the homogeneous gravity field. The disk radius is *a,* height $2h$. Note that for a greater *h* the disk becomes a cylinder and eventually a bar; but the term disk will be used for all cases.

To describe disk motion three coordinate systems are used (Figure 1) ([3],[13]): the inertial coordinate system O*XZY* and the intermediate coordinate systems C*xzy* and C$\xi\eta\zeta$ with origins in the disk centre of mass point C. The system C*xzy* is with respect to O*XZY* rotated about the *Z*-axis at spin angle $\psi$ and the system C$\xi\eta\zeta$ is with respect to C*xzy* rotated about the *x*-axis at inclination angle $\theta$. The body frame of reference (which is not used) is rotated about $\eta$-axis at rotation angle $\varphi$.

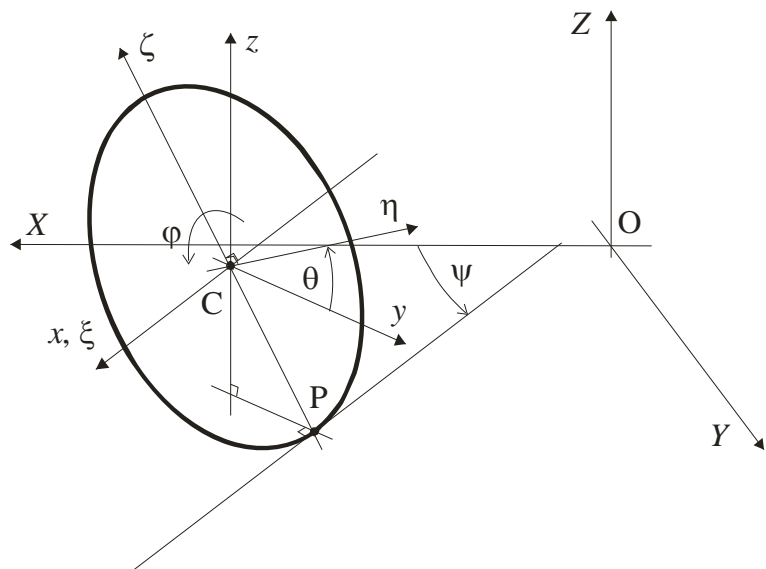

**Figure 1** The coordinate systems used (a thin disk is shown for clarity)

There are three kinds of possible contact of the disk with the plane ([3]). When $\theta = \pm\pi/2$ the disk lies on the plane (surface contact) and this case is excluded from further discussion. When $h > 0$ and $\theta = 0$ the disk contact with the plane is a line on the disk-bounding surface. There are two simple motions of the disk in this position: pure rolling and pure spinning about a diameter. In all other cases when the disk reaches $\theta = 0$ this is accompanied by impact and a bouncing motion can result. These phenomena require additional study, so case $\theta = 0$ will only be used for $h = 0$. For all other angles $\theta$, the disk is inclined to vertical and the contact is point P on the edge between the disk surface and its bounding planes (Figure 2).

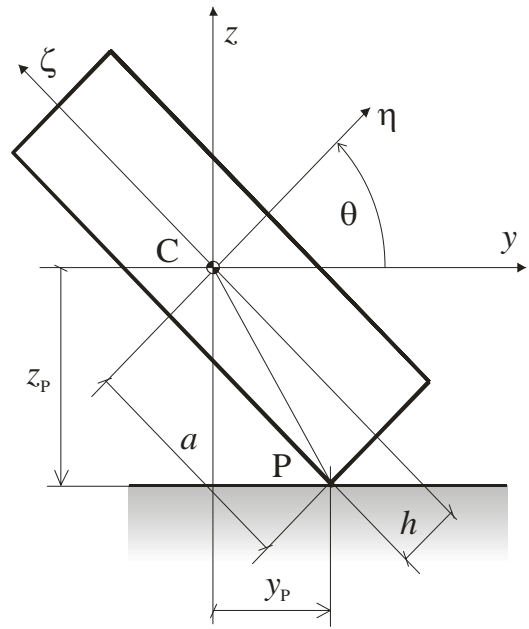

**Figure 2** The position of the contact point

The position of the disk in O$XZY$ is given by coordinates $(X_C, Y_C, Z_C)$ of the point C and it's orientation by angles $(\psi, \theta, \varphi)$. If $(v_{Cx}, v_{Cy}, v_{Cz})$ are components of velocity $\mathbf{v}_C$ of point C in C$xzy$, and $(\omega_1, \omega_2, \omega_3)$ are components of angular velocity $\boldsymbol{\omega}$ with respect to C$\xi\eta\zeta$, then the kinematics of the disk are described by the following sets of differential equations

$$\frac{dX_C}{dt} = v_{Cx}\cos\psi - v_{Cy}\sin\psi \qquad \frac{dY_C}{dt} = v_{Cx}\sin\psi + v_{Cy}\cos\psi \qquad \frac{dZ_C}{dt} = v_{Cz} \qquad (1)$$

$$\frac{d\psi}{dt} = \Omega = \frac{\omega_3}{\cos\theta} \qquad \frac{d\theta}{dt} = \omega_1 \qquad \frac{d\varphi}{dt} = \omega = \omega_2 - \omega_3\tan\theta \qquad (2)$$

where $t$ is time. Here $\Omega$ is the angular velocity of spinning of the system $Cxzy$ around $Z$-axis, and $\omega$ is the angular velocity of rotation of the disk around $\eta$-axis. When they are given, from (2), $\omega_2$ and $\omega_3$ can be calculated from

$$\omega_2 = \omega + \Omega\sin\theta \qquad \omega_3 = \Omega\cos\theta \qquad (3)$$

The coordinates of contact point P are, in system $Cxzy$, given by (Figure 2)

$$y_P = a\sin\theta - \tilde{h}\cos\theta \qquad z_P = -\left(a\cos\theta + \tilde{h}\sin\theta\right) \qquad (4)$$

where $\tilde{h} = h\,\text{sgn}\,\theta$ and in system O$XZY$ they are

$$X_P = X_C - y_P\sin\psi \qquad Y_P = Y_C + y_P\cos\psi \qquad Z_P = Z_C + z_P = 0 \qquad (5)$$

Coordinates $\left(X_P, Y_P\right)$ give the path of the contact point on the plane. The components of contact point velocity $v_P$ are given in $Cxzy$ by

$$v_{Px} = v_{Cx} - a\omega_2 + \tilde{h}\omega_3 \qquad v_{Py} = v_{Cy} - \omega_1 z_P \qquad v_{Pz} = v_{Cz} + \omega_1 y_P = 0 \qquad (6)$$

Note that $(5)_3$ and $(6)_3$ are also the equations expressing conditions for contact between the disk and plane. Thus, in the case of contact,

$$Z_C = -z_P \qquad v_{Cz} = -\omega_1 y_P \qquad (7)$$

Let $(F_x, F_y, F_z)$ be components of reaction force $\boldsymbol{F}$ and $(M_x, M_y, M_z)$ components of reaction moment $\boldsymbol{M}$ at the contact point P in $Cxzy$. The principle of momentum and principle of momentum of momentum gives two sets of equations of motion

$$m\left(\frac{dv_{Cx}}{dt} - v_{Cy}\Omega\right) = F_x \qquad m\left(\frac{dv_{Cy}}{dt} + v_{Cx}\Omega\right) = F_y \qquad m\frac{dv_{Cz}}{dt} = -mg + F_z \qquad (8)$$

$$mk_1^2 \frac{d\omega_1}{dt} = m\left(k_2^2 \omega_2 - k_1^2 \omega_3 \tan\theta\right)\omega_3 - z_P F_y + y_P F_z + M_x$$

$$mk_2^2 \frac{d\omega_2}{dt} = -a\, F_x + M_y \cos\theta + M_z \sin\theta \qquad (9)$$

$$mk_1^2 \frac{d\omega_3}{dt} = m\left(-k_2^2 \omega_2 + k_1^2 \omega_3 \tan\theta\right)\omega_1 + \tilde{h}\, F_x - M_y \sin\theta + M_z \cos\theta$$

where $g$ is gravity acceleration, $m$ mass of disk, and $k_1$ and $k_2$ principal radii of gyration of the disk given by

$$k_1^2 = \frac{a^2}{4} + \frac{h^2}{3} \qquad k_2^2 = \frac{a^2}{2} \qquad (10)$$

Because $v_{Cz}$ is given by $(7)_2$, $(8)_3$ can be used to express the normal reaction force $F_z$. By noting that $\dfrac{dy_P}{dt} = \dfrac{dy_P}{d\theta}\dfrac{d\theta}{dt} = -z_P \omega_1$, one finds

$$F_z = mg + m\frac{dv_{Cz}}{dt} = m\left(g + z_P \omega_1^2 - y_P \frac{d\omega_1}{dt}\right) \qquad (11)$$

The unilateral contact between the disk and the plane is maintained by the condition that the normal reaction is positive; i.e., $F_z > 0$.

From (8) and (9) the following energy equation can be derived

$$\frac{dE}{dt} = \bm{v}_P \cdot \bm{F}_t + \bm{\omega} \cdot \bm{M} \qquad (12)$$

where $\bm{F}_t$ is the friction force where the components are $(F_x, F_y)$ and $E$ is the total mechanical energy of the disk given by

$$E = \frac{m}{2}\left(v_{Cx}^2 + v_{Cy}^2 + v_{Cz}^2\right) + \frac{mk_1^2}{2}\left(\omega_1^2 + \omega_3^2\right) + \frac{mk_2^2}{2}\omega_2^2 + mg\, Z_C \qquad (13)$$

Since energy can not increase, both of the following inequalities must hold

$$\bm{v}_P \cdot \bm{F}_t \leq 0 \qquad \bm{\omega} \cdot \bm{M} \leq 0. \qquad (14)$$

In what follows, the dimensionless form of the equations will be used. They are obtained if units of mass, length and time are $m$, $a$, and $\sqrt{a/g}$, or, equivalently, if one set $m = a = g = 1$.

### 3 Steady State Solutions

Equations (1)$_{1,2}$, (2), (8)$_{1,2}$ and (9) form a set of ten ordinary differential equations with fifteen unknowns

$$X_C, Y_C, \psi, \theta, \varphi, v_{Cx}, v_{Cy}, \omega_1, \omega_2, \omega_3, F_x, F_y, M_x, M_y, M_z$$

which could be (at least numerically) integrated when the initial conditions are specified and the additional five constitutive equations for calculation of friction force $\bm{F}_t$ and contact moment $\bm{M}$, which fulfils inequalities (14), are given. However, in the case of steady motion, which is the interest of this paper, the energy should be constant. Thus from (12) and (14) the steady motion is possible only if $\bm{v}_P \cdot \bm{F}_t = 0$ and $\bm{\omega} \cdot \bm{M} = 0$. The first is true in two cases:

- on perfectly rough ground where slip velocity is zero; i.e., $v_{Px0} = v_{Py0} = 0$ and

- on perfectly smooth ground when friction force is zero; i.e., $F_x = F_y = 0$.

The second is true if $\boldsymbol{M = 0}$ or $\boldsymbol{\omega = 0}$. In what follows only the first is taken to be true since the second implies that the disk is at rest if the ground is perfectly rough or in a state of inertial motion if the ground is perfectly smooth. In what follows the term perfect will be omitted when referring to the ground type.

Besides energy, all unknowns entered into the energy equation should be constants in steady motion as well. According to (13) these unknowns are components of the velocity of point C, components of the angular velocity of the disk and, at $Z_C$, the inclination angle. Let $\theta = \theta_0$ be the steady state value of the inclination angle. Then by $(2)_2$ $\omega_1 = 0$ and further by $(7)_2$ $v_{Cz} = 0$. The remaining unknowns are

$$v_{Cx} = v_{Cx0}, \ v_{Cy} = v_{Cy0} \text{ and}, \omega_2 = \omega_{20}, \ \omega_3 = \omega_{30}$$

where the index 0 is added to variables to denote their constant values in steady state.

The angle $\theta_0$ can be positive or negative but as will be shown later only case $0 \leq \theta_0 < \pi/2$ can be considered. Three possible inclinations of the disk in this range are shown in Figure 3. The position given by $\tan\theta_0 = h$ corresponds to the static equilibrium (SE) position of a disk.

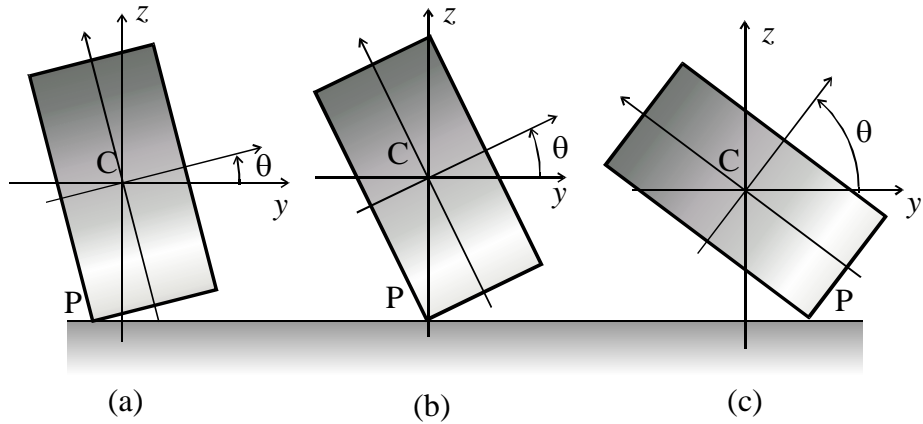

**Figure 3** Three types of possible inclination of the disk for $0 < \theta_0 < \pi/2$.

(a) $\tan\theta_0 < h$, (b) $\tan\theta_0 = h$ - static equilibrium (SE) position, (c) $\tan\theta_0 > h$

Integration of (2)$_{1,3}$ yields

$$\psi = \psi_0 + \Omega_0 t \qquad \varphi = \varphi_0 + \omega_0 t \qquad (15)$$

where $\psi_0, \varphi_0$ are initial angles of $\psi$, $\varphi$ and $\Omega_0$, $\omega_0$ are steady state values of $\Omega$, $\omega$. The coordinates of contact point P in $Cxzy$, are, by (4)

$$y_{P0} = \sin\theta_0 - \tilde{h}\cos\theta_0 \qquad z_{P0} = -\left(\cos\theta_0 + \tilde{h}\sin\theta_0\right) \qquad (16)$$

and in $OXZY$, by (5), are

$$X_{P0} = X_{C0} - y_{P0}\sin(\psi_0 + \Omega_0 t) \qquad Y_{P0} = Y_{C0} + y_{P0}\cos(\psi_0 + \Omega_0 t) \qquad (17)$$

The non-zero components of velocity of point P are from (6)

$$v_{Px0} = v_{Cx0} - \omega_{20} + \tilde{h}\omega_{30} = v_{Cx0} - \omega_0 - y_{P0}\Omega_0 \qquad v_{Py0} = v_{Cy0} \qquad (18)$$

Further, from (8)$_2$, (9)$_2$ and (11) the components of contact force are

$$F_x = 0 \qquad F_y = v_{Cx0}\Omega_0 \qquad F_z = 1 \tag{19}$$

The last equation shows that in the case of steady state, condition $F_z > 0$ is automatically satisfied. Since $\omega_1 = 0$, equation $(9)_3$ is trivially satisfied and the remaining equations of motion $(8)_2$ and $(9)_1$ reduced to the conditions for a steady motion

$$v_{Cy0}\Omega_0 = 0 \qquad \left(k_2^2 \omega_{20} - k_1^2 \omega_{30} \tan\theta_0\right)\omega_{30} - z_{P0}F_y + y_{P0} = 0 \tag{20}$$

Depending on condition $(20)_1$ kinematical equations (1) have two possible solutions. It will be shown that when $\Omega_0 = 0$ a disk rolls on a straight line or stands still, and when $\Omega_0 \neq 0$ the disk rolls in a circle.

### 3.1 Straight line motion

When $\Omega_0 = 0$ it follows from $(15)_1$ that $\psi = \psi_0$, hence (1) integrates to

$$\begin{aligned}
X_C &= X_{C0} + \left(v_{Cx0}\cos\psi_0 - v_{Cy0}\sin\psi_0\right)t \\
Y_C &= Y_{C0} + \left(v_{Cx0}\sin\psi_0 + v_{Cy0}\cos\psi_0\right)t
\end{aligned} \tag{21}$$

where $X_{C0}$, $Y_{C0}$ are the initial position of point C. The solution shows that if $v_{Cx0} \neq 0$ or $v_{Cy0} \neq 0$ then the disk rolls on a straight line and if $v_{Cx0} = v_{Cy0} = 0$ the disk stands at rest. Further, it follows from $(19)_2$ that $F_y = 0$ and since also $F_x = 0$ there is no friction force in the straight line steady motion even in the case of rough ground.

Because when $\Omega_0 = 0$ then also $\omega_{30} = 0$ and the condition $(20)_2$ is reduced to $y_{P0} = 0$. From (17), it follows that $X_P = X_C$, $Y_P = Y_C$; i.e., the coordinates of instantaneous contact point P coincide with those of point C as expected. Also, from $(16)_1$, one has $\tan\theta_0 = \tilde{h}_0$ showing that in straight line steady motion the disk is at the SE position (Figure 3b).

In the case when a disk is rolling on a straight line on rough ground one has $v_{Py0} = v_{Cy0} = 0$ and $v_{Px0} = 0$, which, from (18)$_1$, yields $v_{Cx0} = \omega_0$. Thus on rough ground the only arbitrary parameter influencing the motion is $\omega_0$. On smooth ground $v_{Cx0}$, $v_{Cy0}$ and $\omega_{20}$ are arbitrary and from (18) the components of velocity of point P are given by $v_{Px0} = v_{Cx0} - \omega_0$ and $v_{Py0} = v_{Cy0}$.

### 3.2 Circular motion

When $\Omega_0 \neq 0$ then from (20)$_1$ $v_{Py0} = v_{Cy0} = 0$. Integrals of kinematical equations (1) can then be written in the form

$$X_C = X_A - r_C \sin(\psi_0 + \Omega_0 t) \qquad Y_C = Y_A + r_C \cos(\psi_0 + \Omega_0 t) \qquad (22)$$

and then coordinates of point P, by using (17) and (22), in the form

$$X_P = X_A - r_P \sin(\psi_0 + \Omega_0 t) \qquad Y_P = Y_A + r_P \cos(\psi_0 + \Omega_0 t) \qquad (23)$$

where

$$r_C \equiv -\frac{v_{Cx0}}{\Omega_0} \qquad r_P \equiv r_C + y_{P0} \qquad (24)$$

This shows that points C and P describe concentric circles originating at point $A = (X_A, Y_A)$ with radii $|r_C|$ and $|r_P|$. Note that the coordinates of the initial position of point C at $t = 0$ are, from (22),

$$X_{C0} = X_A - r_C \sin\psi_0 \qquad Y_{C0} = Y_A + r_C \cos\psi_0 \qquad (25)$$

The sign $r_C$ depends on the signs of velocities $v_{Cx0}$ and $\Omega_0$ while the direction of rolling in a circle depends also on sign $\Omega_0$. However, as can be shown from (22) by transforming coordinate system O$XZY$ to point A and rotating it to angle $\psi_0$, that for fixed centre point A, the disk rolls counter clockwise if $\Omega_0 > 0$ and clockwise if $\Omega_0 < 0$, whatever sign $r_C$ is. The sign $r_C$, according to (25), only affects the initial position of point C. From this the relation of signs $r_C$ and inclination angle $\theta_0$ is as follows: if $r_C$ and $\theta_0$ has the same sign then the disk diameter is leaning towards point A; otherwise it leans outward. In other words, negative inclination $\theta_0$ can be replaced by negative $r_C$. This justifies consideration of $\theta_0$ only at interval $0 \leq \theta_0 < \pi/2$ as was stated earlier.

### 3.2.1 Rough Ground

In the case of rough ground, besides $v_{Py0} = 0$ also $v_{Px0} = 0$; thus, from (18)$_1$, $v_{Cx0} = \omega_{20} - \tilde{h}\omega_{30} = \omega_0 + y_{P0}\Omega_0$ or from (24), $v_{Cx0} = -r_C\Omega_0$. Using this (24) can be put into the form

$$r_C = \left(-\frac{\omega_{20}}{\omega_{30}} + \tilde{h}\right)\cos\theta_0 = r_P - y_P \qquad r_P = -\frac{\omega_0}{\Omega_0} \qquad (26)$$

From (19)$_2$ the non-zero friction force component is

$$F_y = -r_C\Omega_0^2 \qquad (27)$$

demonstrating that $F_y$ is directed toward point A. The condition for steady motion (20)$_2$ can be, by using (10) and (16), written in the form

$$\left[\left(\frac{1}{4}+\frac{4}{3}h^2\right)\tan\theta_0+\tilde{h}\right]\omega_{30}^2 - \left(\frac{3}{2}+\tilde{h}\tan\theta_0\right)\omega_{20}\omega_{30} = \sin\theta_0 - \tilde{h}\cos\theta_0 \tag{28}$$

or, in terms of $\Omega_0, \omega_0$, in terms form

$$\begin{aligned}\left[\left(\frac{5}{8}-\frac{2}{3}h^2\right)\sin 2\theta_0 - \tilde{h}\cos 2\theta_0\right]\Omega_0^2 + \left(\frac{3}{2}\cos\theta_0 + \tilde{h}\sin\theta_0\right)\omega_0\Omega_0 \\ + \sin\theta_0 - \tilde{h}\cos\theta_0 = 0\end{aligned} \tag{29}$$

Equations (26), (28) and (29) connect four constants: $\theta_0, \omega_{20}, \omega_{30}, r_C$ or $\theta_0, \Omega_0, \omega_0, r_P$. Two can be chosen arbitrarily; the other two are then determinate. When $\theta_0, \Omega_0$ are given, $\omega_0$ can be calculated from (29) and then $r_C$ and $r_P$ can be calculated from (26). If $\theta_0, r_P$ or $\theta_0, r_C$ are given then the situation is a bit complicated since two cases must be distinguished: the case $\tan\theta_0 = \tilde{h}$ and the case $\tan\theta_0 \neq \tilde{h}$.

First, if $\theta_0, r_C$ are given and $\tan\theta_0 = \tilde{h}$ meaning that the disk is in the SE position, then by (16)$_1$ also $y_P = 0$ and so, from (26), $r_C = r_P$. Now, since in circular motion $\Omega_0 \neq 0$, it follows from (26)$_2$ and (29) that

$$\omega_0 = -r_P\Omega_0 \qquad r_C = r_P = \frac{\tilde{h}}{6}\frac{\left(3-4h^2\right)}{\left(3+2h^2\right)\sqrt{1+h^2}} \tag{30}$$

If $0 \leq \theta_0 < \pi/2$ then, as can be seen from (30), when $h < \sqrt{3}/2$ then $r_C > 0$, indicating that the disk is leaning toward point A and when $h > \sqrt{3}/2$ then $r_C > 0$ so the disk leans outward. The radius $r_C$ is in both cases limited. When $h < \sqrt{3}/2$ the maximal value of $r_C$ is for $h = \sqrt{33\sqrt{93}-231}/22 \approx 0.425$ at $\theta_0 \approx 23^0$ where $r_C \approx 0.0442$. When $h \to \infty$, implying also $\theta_0 \to \pi/2$, then the limit is $r_C \to -\frac{1}{3}$.

When $r_C = r_P = 0$ --i.e., points C and P are at rest--the disk spins in place with $\Omega_0$ and according to (26), $\omega_0 = 0$. From (30) this can be achieved when $h = 0$ or when $h = \sqrt{3}/2 \approx 0.866$. The corresponding inclination angle in the last case is $\theta_0 = \arctan \sqrt{3}/2 \approx 0.714 \approx 40.893^0$.

In the second case, when $\theta_0, r_C$ are given and $\tan \theta_0 \neq \tilde{h}$, from (26) and (29) one obtains

$$\Omega_0^2 = \frac{\tan \theta_0 - \tilde{h}}{\cos \theta_0 \left[ \left( \frac{1}{4} + \frac{h^2}{3} \right) \tan \theta_0 - \tilde{h} \right] + \left( \frac{3}{2} + \tilde{h} \tan \theta_0 \right) r_C}$$
$$= \frac{\sin \theta_0 - \tilde{h} \cos \theta_0}{\left( \frac{3}{2} \cos \theta_0 + \tilde{h} \sin \theta_0 \right) r_P - \left( \frac{5}{8} - \frac{2}{3} h^2 \right) \sin 2\theta_0 + \tilde{h} \cos 2\theta_0} \quad (31)$$

For $h = 0$ this equation agrees with those of McDonald ([8], eq 29) noting that their $\alpha_0$ corresponds to the present $\pi/2 - \theta_0$. For steady motion the condition $\Omega_0^2 > 0$ must be fulfilled. Thus, the steady state motion for the given $r_C$ is possible if the numerator and denominator of (31) have the same sign. Let

$$r_C^* = -\cos \theta_0 \frac{\left[ \left( 3 + 4h^2 \right) \tan \theta_0 - 6h \right]}{6 \left( 3 + 2h \tan \theta_0 \right)} \quad (32)$$

be the radius where the denominator of (31) becomes zero, then for $\tan \theta_0 < h$ possible radii are limited upward: i.e., $-\infty < r_C < r_C^*$; and for $\tan \theta_0 > h$ the radii are limited downward: i.e., $r_C^* < r_C < \infty$. Also, for a given $h$, radius $r_C^*$ is bounded. The maximum value of $r_C^*$ is for $\theta_0 = 0$ where $r_C^* = \frac{h}{3}$, and the minimum value is between zeros of $r_C^*$ which are at $\theta_0 = \arctan \frac{6h}{3 + 4h^2}$ and $\theta_0 = \pi/2$, except in the case $h = 0$ when the minimum is at $\theta_0 = \pi/2$ where $r_C^* = -1/6$ (Figure 4a). For a great $h$, the minimum value between the zeros of $r_C^*$ tends towards $r_C^* \to -h/3$. Examples of graphs of $\Omega_0^2 > 0$ for various $h$ are shown in Figure 4 where dark regions are those where $\Omega_0^2 < 0$ and $r_C^*$ is the boundary between white and dark regions. Gray regions belong to possible but unstable steady motions. The explanation of this is postponed to the next section.

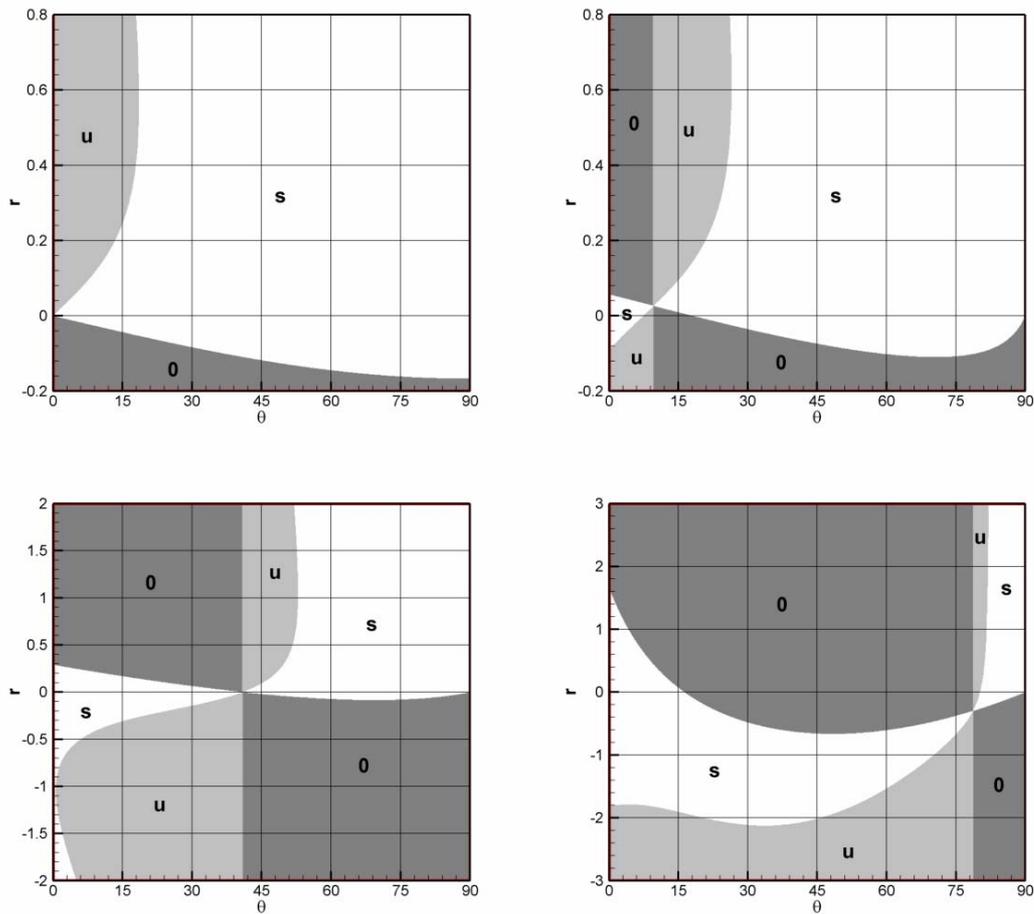

**Figure 4.** Graphs of $\Omega_0^2 > 0$ with the correspondent stability regions $\varpi^2 > 0$.
for $h = 0$, $h = 0.17$, $h = \sqrt{3}/2$ and $h = 5$.
(s – stable, u – unstable, 0 – indefinite)

When $r_C = 0$, point C is at rest. In this case, from (26), $r_P = y_{P0}$ and $\omega_0 = -y_P \Omega_0$ and from (27), $F_y = 0$; i.e., there is no friction force in such a motion. The condition for steady motion (31) becomes

$$\Omega_0^2 = 12 \frac{\tan\theta_0 - \tilde{h}}{(3 + 4h^2)\sin\theta_0 - 12\tilde{h}\cos\theta_0} \qquad (33)$$

Now, $\Omega_0^2 > 0$ will be satisfied if for $\tan\theta_0 < h$ also $\tan\theta_0 < \dfrac{12h}{3+4h^2}$ and for $\tan\theta_0 > h$ also $\tan\theta_0 > \dfrac{12h}{3+4h^2}$. This implies that for $h < \sqrt{3}/2$ steady motion is possible if $\tan\theta_0 < h$ or $\tan\theta_0 > \dfrac{6h}{3+4h^2}$ and for $h > \sqrt{3}/2$ if $\tan\theta_0 > h$ or $\tan\theta_0 < \dfrac{6h}{3+4h^2}$. As $\theta_0 \to \pi/2$, the spinning angular velocity becomes very great since in this case one has asymptotically $\Omega_0^2 \sim \dfrac{12}{(3+4h^2)} \tan\theta_0$. On the other hand, when $r_C \neq 0$, then, for fixed $h > 0$, as $\theta_0 \to \pi/2$, the limit is $\Omega_0^2 \to \dfrac{1}{hr_C}$. Again for small $r_C$ or $h$ the spinning angular velocity becomes very great.

### *3.1.2 Smooth Ground*

On smooth ground $F_y = 0$, so from (19)$_2$, $v_{Cx0} = 0$ and then from (24)$_1$ also $r_C = 0$. This shows that point C is at rest and that contact point P, as follows from (24)$_2$, describes the circle with radius $r_P = y_{P0}$. The non-zero component of velocity of contact point P is, according to (18), $v_{Px0} = -\omega_{20} + \tilde{h}\omega_{30} = -(\omega_0 + y_{P0}\Omega_0)$, thus the condition for steady motion (20)$_2$ becomes

$$\left(\frac{1}{4} + \frac{h^2}{3}\right)\tan\theta_0 \omega_{30}^2 - \frac{1}{2}\omega_{20}\omega_{30} = \sin\theta_0 - \tilde{h}\cos\theta_0 \qquad (34)$$

or in terms of $\Omega_0, \omega_0$, by using (3),

$$\left(\frac{1}{4} - \frac{h^2}{3}\right)\Omega_0^2 \sin\theta_0 + \frac{1}{2}\omega_0\Omega_0 + \tan\theta_0 - \tilde{h} = 0 \qquad (35)$$

For $h = 0$, formula (35) agrees with those given by Milne ([9]) noting that his $\alpha$ corresponds to the present $\theta_0 - \pi/2$ and his $n$ to $-\omega_0$. In particular, when $h = \tan\theta_0$; i.e., when the disk is in the SE position, then also $r_P = 0$ so the disk spins in place with $\Omega_0$. In this case $\omega_0 = -\left(\dfrac{1}{2} - \dfrac{2h^2}{3}\right)\Omega_0 \sin\theta_0$.

## 4 Small Oscillations about Steady Motion

The solution for small oscillations about steady motion will be obtained in the usual way by assuming the solution of system (8)$_{1,2}$ and (9) in the form

$$\theta = \theta_0 + \tilde{\theta} \qquad \omega_1 = \tilde{\omega}_1 = \frac{d\tilde{\theta}}{dt} \qquad \omega_2 = \omega_{20} + \tilde{\omega}_2 \qquad \omega_3 = \omega_{30} + \tilde{\omega}_3 \qquad (36)$$

where $\tilde{\theta}$, $\tilde{\omega}_1$, $\tilde{\omega}_2$, $\tilde{\omega}_3$ are small perturbations of corresponding variables in the stationary state. Unlike the steady motion solution the case of rough ground and smooth ground must be considered separately.

### *4.1 Rough Ground*

On rough ground the instantaneous contact point P is at rest; i.e., $v_{Px} = v_{Py} = v_{Pz} = 0$. Using this, the velocity components $v_{Cx}, v_{Cy}$ of point C are given, from (6)$_{1,2}$, by

$$v_{Cx} = \omega_2 - \tilde{h}\omega_3 \qquad v_{Cy} = -\omega_1\left(a\cos\theta + \tilde{h}\sin\theta\right) \qquad (37)$$

When these are substituted into (8)$_{1,2}$ the expressions for components of friction force $F_x, F_y$ are obtained in terms of inclination angle $\theta$ and components of angular velocities $\omega_1, \omega_2$ and $\omega_2$. When these expressions and $F_z$, which is given by (11), are inserted into (9) the following system of equations is obtained

$$\begin{aligned}
\frac{d\omega_1}{dt} &= \frac{1}{(15+16h^2)}\Big\{6\big(3+2\tilde{h}\tan\theta\big)\omega_2\omega_3 \\
&\quad -\big[(3+16h^2)\tan\theta+12\tilde{h}\big]\omega_3^2+12\big(\sin\theta-\tilde{h}\cos\theta\big)\Big\} \\
\frac{d\omega_2}{dt} &= -\frac{1}{(3+8h^2)}\bigg[4\tilde{h}\omega_2+\frac{2}{3}(3+4h^2)\omega_3\bigg]\omega_1 \\
\frac{d\omega_3}{dt} &= \bigg[-\frac{6}{3+8h^2}\omega_2+\bigg(\tan\theta+\frac{4\tilde{h}}{3+8h^2}\bigg)\omega_3\bigg]\omega_1
\end{aligned} \qquad (38)$$

For $h=0$ these equations agree with those of O'Reilly ([13]). When (36) is substituted into (38) and higher order terms are neglected one obtains

$$\begin{aligned}
\frac{d^2\tilde{\theta}}{dt^2}+\varpi^2\tilde{\theta} &= 0 \\
\frac{d\tilde{\omega}_2}{dt} &= -\frac{12\tilde{h}\omega_{20}+2(3+4h^2)\omega_{30}}{3(3+8h^2)}\tilde{\omega}_1 \\
\frac{d\tilde{\omega}_3}{dt} &= \bigg[\frac{-6\omega_{20}}{(3+8h^2)}+\bigg(\tan\theta_0+\frac{4\tilde{h}}{3+8h^2}\bigg)\omega_{30}\bigg]\tilde{\omega}_1
\end{aligned} \qquad (39)$$

Here $\varpi$ is the angular frequency of steady oscillation and is given by

$$\varpi^2 = \frac{\big(P_{22}\omega_{20}^2+P_{23}\omega_{20}\omega_{30}+P_{33}\omega_{30}^2\big)\sqrt{1+p_0^2}-12(3+8h^2)(1+\tilde{h}p_0)}{(3+8h^2)(15+16h^2)\sqrt{1+p_0^2}} \qquad (40)$$

where

$$\begin{aligned}
P_{22} &\equiv 36(3+2\tilde{h}p_0) \\
P_{23} &\equiv -6\big[4\tilde{h}(3+8h^2)p_0^2+(15+56h^2)p_0+2\tilde{h}(15+8h^2)\big] \\
P_{33} &\equiv 3(3+16h^2)(3+8h^2)p_0^2+8\tilde{h}(15+44h^2)p_0+45+216h^2+128h^4
\end{aligned} \qquad (41)$$

and $p_0=\tan\theta_0$. For $h=0$ (39) agrees with those given in ([12]).

The motion is stable if $\varpi^2 > 0$. For straight line motion, discussed in section 3.1, $\omega_{30} = 0$ and $p_0 = \tilde{h}$, therefore from (40) the rolling will be stable if

$$\omega_{20}^2 > \frac{(3+8h^2)\sqrt{1+h^2}}{3(3+2h^2)} \tag{42}$$

For $h=0$ one has $\omega_{20}^2 > \frac{1}{3}$, which agrees with values given by other authors ([8],[13],[14]). In the case of circular motion, discussed in section 3.2.1, two cases were distinguished. When $p_0 = \tilde{h}$ and $r_C = r_P$ the motion will be stable if

$$\omega_{30}^2 > \frac{12\sqrt{1+h^2}}{(1+h^2)(15+16h^2)} \tag{43}$$

Again, for $h=0$ (43) gives $\omega_{30}^2 > \frac{4}{5}$, which also agrees with values given by other authors ([8],[13],[14]). For spinning case $r_C = r_P = 0$ when $h = \sqrt{3}/2$ (43) yields $|\omega_{30}| > \sqrt{\frac{8\sqrt{7}}{63}} = 0.580$. When $p_0 \neq \tilde{h}$, $\omega_{20}$ can be expressed from (28) as

$$\omega_{20} = \frac{\left[(3+4h^2)p_0 + 12\tilde{h}\right]}{6(3+2\tilde{h}p_0)}\omega_{30} - 2\frac{p_0 - \tilde{h}}{(3+2\tilde{h}p_0)\sqrt{1+p_0^2}}\frac{1}{\omega_{30}} \tag{44}$$

and when this is substituted into (40) the following expression is obtained

$$\varpi^2 \equiv \frac{P_2\omega_{30}^4 + P_1\omega_{30}^2 + P_0}{(3+8h^2)(15+16h^2)(1+p_0^2)^{3/2}(3+2\tilde{h}\,p_0)\omega_{30}^2} \tag{45}$$

where

$$P_0 \equiv 144\left(p_0 - \tilde{h}\right)^2$$

$$P_1 \equiv -12\left(3 + 8h^2\right)\left(1 + p_0^2\right)\left(3 + 2h^2 + 6\tilde{h}p_0 - \left(3 - 6h^2\right)p_0^2 - 4\tilde{h}p_0^3\right)$$

$$P_2 \equiv \left(3 + 8h^2\right)\left(1 + p_0^2\right)^{3/2}\left[45 + 24h^2 + 84\tilde{h}p_0 + \left(15 + 104h^2\right)p_0^2 + 2\tilde{h}\left(3 + 16h^2\right)p_0^3\right]$$

Since the denominator of (45) is always positive, stability is controlled by its numerator. Thus, for stability

$$P_2\omega_{30}^4 + P_1\omega_{30}^2 + P_0 > 0 \tag{46}$$

Because $P_0 \geq 0$ and $P_2 > 0$ this inequality holds for any value of $\omega_{30}$ if $P_1 > 0$ or discriminant of (46) is negative; i.e., $\Delta \equiv P_1^2 - 4P_0P_2 < 0$ ([5]). The graphs of these inequalities are shown in Figure 5. It turns out that the stability boundaries result from condition $\Delta < 0$. For great $h$ and $p_0$ --i.e., when $\theta_0 \to \pi/2$ --this condition becomes $\theta_0 > \arctan\dfrac{3h}{2}$. For $h = 0$ (46) is reduced to $\tan\theta_0 > \dfrac{\sqrt{198\sqrt{5} - 429}}{11}$, which gives the value $\theta_0 \approx 0.325 = 18.6^0$. This value agrees with that given by Kuleshov ([5]).

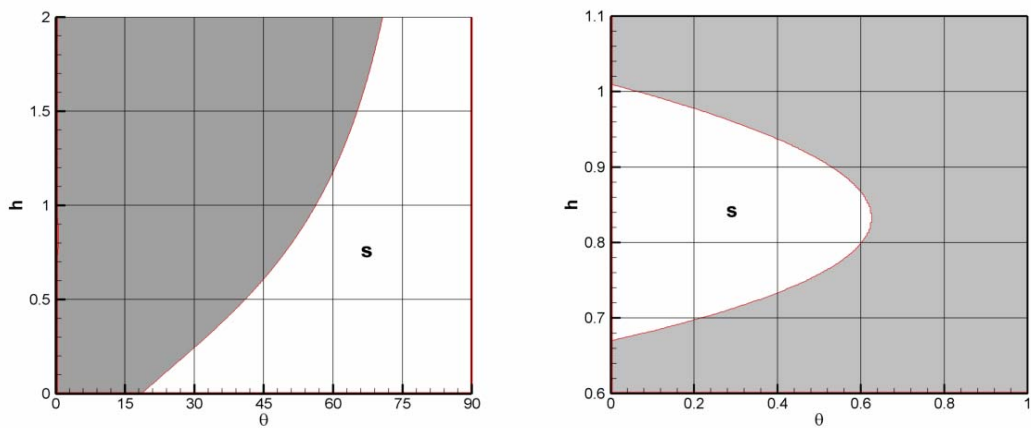

**Figure 5.** Graph of $P_1 > 0 \cup \Delta < 0$. The right graph magnifies the small stable region near $\theta_0 = 0$. In regions denoted by 's' the motion is stable for any value of $\omega_{30}$. Values of $\theta$ are in degrees.

If $\omega_{30}$ is expressed in terms of $\Omega_0$, then by using (31) the angular frequency $\varpi$ can be expressed in terms of $r_C$ in the form

$$\varpi^2 = 12 \frac{R_2 r_C^2 + R_1 r_C + R_0}{Q_1 r_C + Q_0} \tag{47}$$

where $R_2, R_1, R_0$ and $Q_1, Q_0$ are polynomials of variables $p_0, h$. Note that the stability condition $\varpi^2 > 0$ expressed in the form (47) must be considered together with condition $\Omega_0^2 > 0$ since otherwise (47) predicates some non-existing stable regions where both $\Omega_0^2 < 0$ and $\varpi^2 < 0$. Examples of the stability regions given by condition $\varpi^2 > 0$ for various $h$ are shown in Figure 4. It is seen from these graphs that for a given $h$ when inclination is $\theta_0 < \arctan h$ and the disk is leaning toward centre point A then $r_C < r_C^*$ and the motion is stable. When the disk is leaning away from point A the radius $r_C$ is unlimited but the motion becomes unstable with a larger $r_C$. When $\theta_0 > \arctan h$ and the disk is leaning toward point A the radius $r_C$ is unlimited and the motion is stable except for the regions near $\theta_0 = \arctan h$, which become narrow with a larger $r_C$. When the disk is leaning toward point A then $r_C > r_C^*$ and the motion is stable.

The above conditions provide stability regions for any value of $\omega_{30}$. Regions classified as unstable require additional discussion. In Figure 6 the surface $\varpi^2 = 0$ in $(\theta_0, h, \omega_{30})$ space is shown. Note that the graph in Figure 5 is the shadow of that surface projected orthogonally on $(\theta_0, h)$ plane. It is seen that $\varpi^2 = 0$ is limited in the direction of $\omega_{30}$. Its greatest values are for $\theta_0 = 0$. When this is the case one has for $h = 0$ the value $\omega_{30} = \sqrt{4/5}$ and for $h \to \infty$ the limit $\varpi^2 \to \frac{1}{2}(\omega_{30}^2 - 1)$. From this it follows that the motion is stable for any $h > 0$ and $0 \leq \theta_0 < \pi/2$ if $\omega_{30}^2 > 1$.

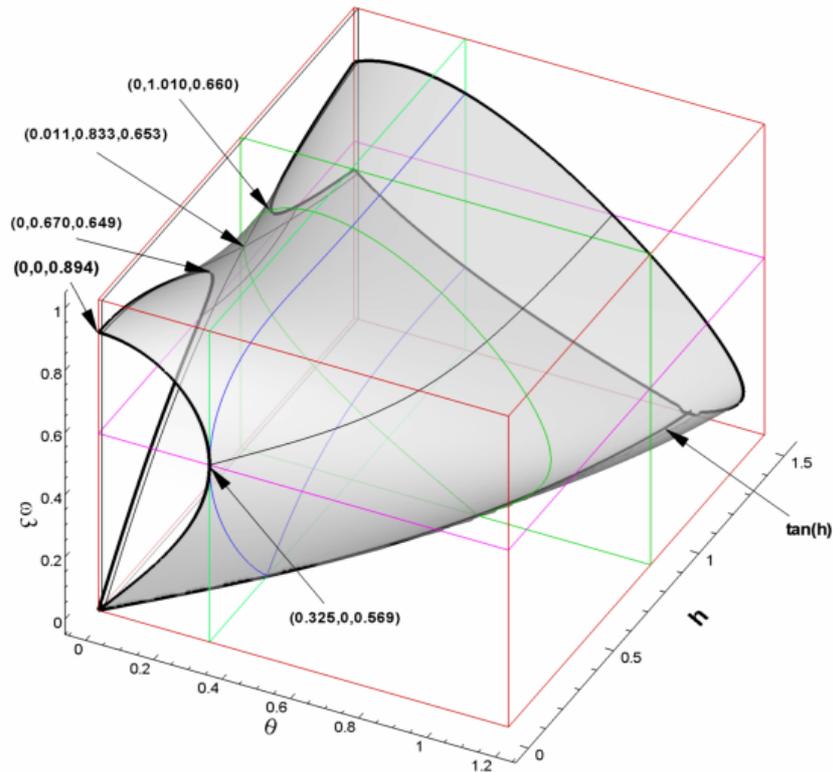

**Figure 6** Surface $\varpi^2 = 0$ given by (45). For values $(\theta_0, h, \omega_{30})$ outside the surface the steady motions are stable. Values of $\theta$ are in radians.

The discussion of the circular motion is concluded by the observation that at limit state as $\theta_0 \to \pi/2$ the angular frequency of steady oscillation becomes very large since in this case, by (45), asymptotically

$$\varpi^2 \sim 2\omega_{30}^2 \frac{(3+16h^2)}{(15+16h^2)} \tan^2 \theta_0 \tag{48}$$

### *4. 2 Smooth Ground*

On smooth ground the friction force is zero $F_x = F_y = 0$. Hence, by using $F_z$ from (11), (9) becomes

$$\left(k_1^2 + y_P^2\right)\frac{d\omega_1}{dt} = \left(k_2^2\omega_2 - k_1^2\omega_3 \tan\theta\right)\omega_3 + y_P + y_P z_P \omega_1^2$$

$$\frac{d\omega_2}{dt} = 0 \qquad (49)$$

$$k_1^2 \frac{d\omega_3}{dt} = \left(-k_2^2 \omega_2 + k_1^2 \omega_3 \tan\theta\right)\omega_1$$

When (36) is substituted into (49) and higher order terms are neglected it becomes

$$\frac{d^2\tilde\theta}{dt^2} + \varpi^2 \tilde\theta = 0 \qquad \frac{d\tilde\omega_2}{dt} = 0 \qquad \frac{d\tilde\omega_3}{dt} = \left(\frac{-6\omega_{20}}{3+4h^2} + \tan\theta_0\,\omega_{30}\right)\tilde\omega_1 \qquad (50)$$

where $\varpi$ is the angular frequency of steady oscillation given by

$$\varpi^2 = \frac{\left(1+p_0^2\right)\left[36\omega_{20}^2 - 18\left(3+4h^2\right)p_0\omega_{20}\omega_{30} + \left(3+4h^2\right)^2\left(1+3p_0^2\right)\omega_{30}^2\right]}{\left(3+4h^2\right)\left[3+16h^2 - 24\tilde{h}p_0 + \left(15+4h^2\right)p_0^2\right]}$$
$$- \frac{12\left(1+\tilde{h}p_0\right)\sqrt{1+p_0^2}}{\left[3+16h^2 - 24\tilde{h}p_0 + \left(15+4h^2\right)p_0^2\right]} \qquad (51)$$

For a straight line, from paragraph 3.1, $\omega_{30} = 0$ and $p_0 = \tilde{h}$ so (51) is reduced to

$$\varpi^2 = \frac{36}{\left(3+4h^2\right)^2}\omega_{20}^2 - \frac{12\sqrt{1+h^2}}{3+4h^2} \qquad (52)$$

For stable motion, $\varpi^2 > 0$. This condition is satisfied if

$$\omega_{20}^2 > \frac{1}{3}\left(3+4h^2\right)\sqrt{1+h^2} \qquad (53)$$

For $h = 0$ we have $\omega_{20}^2 > 1$ which comports with O'Reilly ([13]). In the case of circular motion $\omega_{30} \neq 0$ so, from (34),

$$\omega_{20} = \left(\frac{1}{2} + \frac{2h^2}{3}\right) p_0 \omega_{30} - 2 \frac{p_0 - \tilde{h}}{\sqrt{1+p_0^2}} \frac{1}{\omega_{30}} \qquad (54)$$

If this is introduced into (51) one finds

$$\varpi^2 \equiv \frac{P_2 \omega_{30}^4 + P_1 \omega_{30}^2 + P_0}{\left(3+4h^2\right)\left[3+16h^2 - 24\tilde{h} p_0 + \left(15+4h^2\right) p_0^2\right] \omega_{30}^2} \qquad (55)$$

where

$$\begin{aligned} P_0 &\equiv 144\left(p_0 - \tilde{h}\right)^2 \\ P_1 &\equiv -12\left(3+4h^2\right)\left(1+2\tilde{h}p_0 - p_0^2\right)\sqrt{1+p_0^2} \\ P_2 &\equiv \left(3+4h^2\right)^2 \left(1+p_0^2\right)^2 \end{aligned} \qquad (56)$$

For a disk spinning in place, $p_0 = \tilde{h}$, and in this case, according to (55), the motion will be stable if

$$\omega_{30}^2 > \frac{12}{\left(3+4h^2\right)\sqrt{1+h^2}} \qquad (57)$$

Again for $h=0$ one has $\omega_{30}^2 > 4$, which agrees with ([13]). For $p_0 \neq \tilde{h}$, as can be demonstrated, the denominator of (55) has no zeros, so stability is controlled by the numerator

$$P_2 \omega_{30}^4 + P_1 \omega_{30}^2 + P_0 > 0 \qquad (58)$$

According to (56), $P_0 > 0$ and $P_2 > 0$, therefore this inequality holds for any value of $\omega_{30}$ if $P_1 > 0$ or its discriminant is negative; i.e., $\Delta \equiv P_1^2 - 4P_0 P_2 < 0$. The stability regions implied by those conditions are shown in Figure 7.

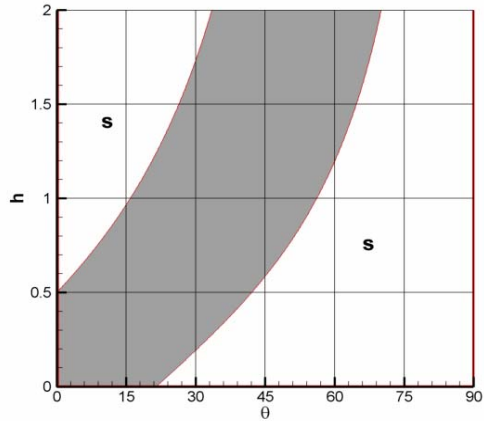

**Figure 7** Graph of $P_1 > 0 \cup \Delta < 0$ for a disk on smooth ground.

In regions marked by 's' the motion is stable for any $\omega_{30}$.

Values of $\theta$ are in degrees.

It turns out that the stability regions depend only on condition $\Delta < 0$, which further reduces to the condition

$$3p_0^4 - 4hp_0^3 + 6p_0^2 - 12\tilde{h}p_0 - 1 + 4h^2 > 0 \qquad (59)$$

For a greater $h$ and $p_0$ the asymptotic of the right limit of (59) is $\theta_0 > \arctan\dfrac{4h}{3}$. In the special case $h = 0$ the motion is stable for any $\omega_{30}$ if $\tan\theta_0 > \dfrac{\sqrt{6\sqrt{3}-9}}{3}$, which gives the angle $\theta_0 \approx 0.375 \approx 21.5^0$. At this angle the limit value is $\omega_{30} \approx 1.167$.

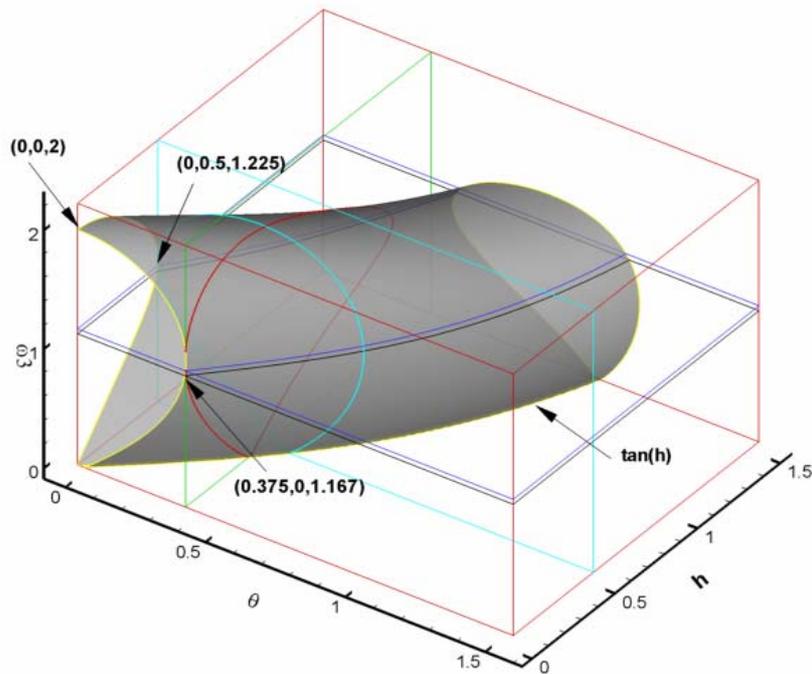

**Figure 8** Plot of $\varpi^2 = 0$ given by (55). For values $(\theta_0, h, \omega_{30})$ outside the shaded surface the motion is stable. Values of $\theta$ are in radians.

As in the case of rough ground, regions classified as unstable require additional discussion. Figure 8 shows the surface $\varpi^2 = 0$, given by (55), in $(\theta_0, h, \omega_{30})$ space. It is seen that the surface is limited in the direction of $\omega_{30}$ and has greatest values for $\theta_0 = 0$ where for $h = 0$ value $\omega_{30} = 2$. For $\theta_0 = 0$ the surface is limited also in $h$ direction where the limit value is $h = 1/2$ where $\omega_{30} = \sqrt{3/2} \approx 1.225$. As $h \to \infty$ then as follows from (57), $\omega_{30}^2 \to 0^+$. From this the conclusion is reached that if $\omega_{30}^2 > 2$ the motion is stable for any $h$ and $\theta_0$.

## Conclusion

The article provides a complete study of steady motion and the stability of steady motion for a disk of finite thickness on rough and smooth surfaces. It was shown that the derived formulas for the condition of steady motion and stability of such a motion in limit cases agrees with those known for an infinitely thin disk. In addition, it was shown that steady motion is stable for any $h > 0$ and $0 \leq \theta_0 < \pi/2$ if in the case of rough ground $\omega_{30}^2 > 1$ or in the case of smooth ground $\omega_{30}^2 > 2$.